\documentclass[sigconf, noacm]{acmart}

\copyrightyear{2025}
\acmYear{2025}
\setcopyright{cc}
\setcctype{by-nc}
\acmConference[CHI '25]{CHI Conference on Human Factors in Computing Systems}{April 26-May 1, 2025}{Yokohama, Japan}
\acmBooktitle{CHI Conference on Human Factors in Computing Systems (CHI '25), April 26-May 1, 2025, Yokohama, Japan}\acmDOI{10.1145/3706598.3713448}
\acmISBN{979-8-4007-1394-1/25/04}

\usepackage{hyperref}
\usepackage{xcolor}
\usepackage{enumitem}
\usepackage{soul}
\usepackage{algorithm}
\usepackage{algpseudocode}

\usepackage{ljhnick}
\usepackage{tabularx}
\usepackage{tabularray}
\usepackage{alltt}
\usepackage{listings}

\sethlcolor{blue!20}

\newcolumntype{C}[1]{>{\centering\let\newline\\\arraybackslash\hspace{0pt}}m{#1}}

\newenvironment{colortext}[1][black]{\color{#1}}{}
\newenvironment{texthighlight}[1][black]{\color{#1}}{}

\begin{document}







\title{\codename: Contextually Augmenting Captured Multimodal Memory to Enable Personal Question Answering}


\author{Jiahao Nick Li}
\affiliation{%
  \institution{UCLA}
  \city{Los Angeles}
  \country{USA}}
\email{ljhnick@ucla.edu}

\author{Zhuohao (Jerry) Zhang}
\affiliation{%
  \institution{University of Washington}
  \city{Seattle}
  \country{USA}}
\email{zhuohao@uw.edu}

\author{Jiaju Ma}
\affiliation{%
  \institution{Stanford University}
  \city{Palo Alto}
  \country{USA}}
\email{jiajuma@stanford.edu}

\renewcommand{\shortauthors}{Li et al.}

\begin{abstract}


People often capture memories through photos, screenshots, and videos. 
While existing AI-based tools enable querying this data using natural language, they only support retrieving individual pieces of information like certain objects in photos, and struggle with answering more complex queries that involve interpreting interconnected memories like sequential events. 
We conducted a one-month diary study to collect realistic user queries and generated a taxonomy of necessary contextual information for integrating with captured memories.
We then introduce \codename, a novel system that is able to answer complex personal memory-related questions that require extracting and inferring contextual information.
\codename augments individual captured memories through integrating scattered contextual information from multiple interconnected memories.
Given a question, \codename retrieves relevant augmented memories and uses a large language model (LLM) to generate answers with references.
In human evaluations, we show the effectiveness of \codename with an accuracy of 71.5\%,
outperforming a conventional RAG system by winning or tying for 74.5\% of the time.
\end{abstract}


\received{20 February 2007}
\received[revised]{12 March 2009}
\received[accepted]{5 June 2009}

\begin{teaserfigure}
    \centering
    \includegraphics[width=1\linewidth]{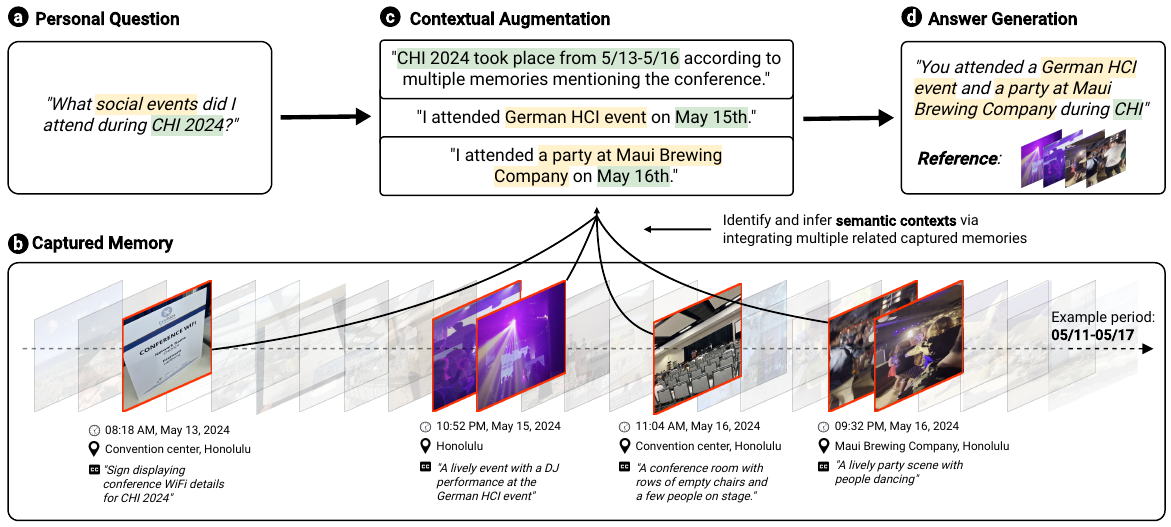}
    \caption{
    \codename is able to answer complex personal questions~(a) on individuals' captured memories~(b), such as captured photos, saved screenshots, and recorded videos. 
    It augments the captured memories by identifying and integrating contextual information scattered across multiple interconnected memories (c).
    \codename then uses this information to retrieve relevant memories and leverages an LLM to generate a comprehensive answer with reference memories (d).}
    \label{fig:teaser}
\end{teaserfigure}

\maketitle

\section{INTRODUCTION}




People often record their everyday life by taking photos, screenshots, and videos for 
saving important information, documenting special occasions, or simply capturing a funny moment~\cite{omniactions}.
These recorded instances, referred to as \textbf{captured memories}, collectively represent subsets of an individual's \textit{episodic memories}~\cite{tulving2002episodic}, a type of long-term memory that contains both specific past experiences and associated contextual details. 
These episodic memories are essential for answering higher-level memory-related personal questions like "\textit{What social events did I attend during CHI 2024?}" (Figure~\ref{fig:teaser}a).
Being able to do so could help users reflect on past experiences and make informed decisions in daily tasks.


However, these raw captured memories by themselves are insufficient to answer personal questions, as they lack contextual details that are typically implicit and scattered across multiple pieces of data.
As shown in Figure~\ref{fig:teaser}b,
memories of attending parties during CHI 2024 are not explicitly annotated as occurring during the event.
Answering such personal questions requires extracting and integrating contextual information not contained within a single captured instance.
For example, by integrating multiple memories that mention ``CHI 2024'' in their content and extracting their metadata, it is possible to determine when the users attended the conference and connect related social events memories from that period to CHI 2024 (Figure~\ref{fig:teaser}c), enabling the answering of the query (Figure~\ref{fig:teaser}d).

Advancements in AI have enable question answering (QA) on long documents~\cite{beltagy2020longformer, wang2024novelqa}, knowledge graph~\cite{huang2019knowledge, yasunaga2021qa}, multimodal databases~\cite{talmor2021multimodalqacomplexquestionanswering, Chen2022MuRAGMR}, and egocentric videos~\cite{mangalam2023egoschemadiagnosticbenchmarklongform, hou2023groundnlqego4dnatural}.
These methods typically rely on data-driven approaches to train powerful models for the target task.
However, the private nature of captured memories makes it difficult to curate large datasets, posing challenges for training models specifically for QA on personal data.
Recent LLM-based work has adopted retrieval augmented generation (RAG) workflows to handle external databases without specific training~\cite{lewis2021retrievalaugmentedgenerationknowledgeintensivenlp}. 
However, such methods depend on explicit connections between queries and relevant external data~\cite{edge2024localglobalgraphrag}. 
In contrast, captured memories are often unstructured and lack contextual annotations, making it difficult to establish explicit links between queries and scattered memories.

To facilitate QA on personal captured memories, we propose \codename, a novel approach designed to robustly and comprehensively answer users' queries on their captured memories. 
\codename has two key components:
\one a question-agnostic pipeline to augment captured memories with contextual information extracted from other related memories to produce \textit{context-augmented memories}, 
and \two a natural language QA system that retrieves these processed memories and generates comprehensive answers with referenced captured memories as evidence.
%
The design of \codename{} is informed by a taxonomy of contextual information that we generated from a one-month diary study with 29 participants.
Specifically, we collected and analyzed 299 user queries
to identify three types of personal questions (direct content queries, contextual filters, and hybrid queries) and three categories of contextual information (atomic context, composite context, and semantic knowledge).
For~\two, \codename employs a retrieval-augmented architecture: given a user input query, it
augments the query via a rewriting strategy, retrieves related memories from the augmented data, and generates the final answer with referenced memory instances via an LLM.


To evaluate \codename, we conducted a user evaluation with 10 participants against a generic RAG-based baseline.
The participants tested queries both logged during the diary study and generated during the evaluation session on a subset of their own captured memories. 
For each tested query, participants rated the user perceived correctness and completeness of the answers generated by both systems in a blinded manner.
The results show that \codename effectively answers different types of queries on users' personal memories, outperforming the baseline with higher accuracy (71.5\%, exceeding the baseline by 27.6\%) and winning or tying 74.5\% of the time in direct comparisons.

\vspace{2mm}
\noindent
In summary, we contribute:
\begin{itemize}[leftmargin=3mm]
    \item A taxonomy of contextual information for augmenting captured memories, derived from queries collected in a one-month diary study with 29 participants.
    

    \item A taxonomy-based pipeline of augmenting captured memories that leverages temporal-based reasoning to extract and infer missing contextual information from other related memories.

    \item The design and implementation of an end-to-end taxonomy-informed system for personal QA\footnote{\codename is open-source at: \url{https://github.com/ljhnick/omniquery}} .

    \item 
    A user evaluation of \codename against a baseline system, demonstrating \codename's effectiveness with 71.5\% accuracy and outperforming the baseline (winning or tying 74.5\% of the time).
    
\end{itemize}

\section{Related Work}

\subsection{Personal Memory Augmentation}
A large body of work in human-computer interaction (HCI) has explored how to augment users' memories. 
This includes developing reminder tools for elderlies or people with memory impairments~\cite{Caprani2006ARO, jamieson2014efficacy, Jamieson2017ForgetMeNotAR, Shin2021PENCODERDF}, providing proactive support in daily tasks~\cite{chi2024memoro, chan2022augmenting}, or manipulating users' memory focus in extended reality~\cite{Bonnail2023MemoryMI}.
These works typically focus on the ``capturing'' stage of the memory augmentation, where researchers develop wearable devices that continuously capture data using designated sensors, which record various modalities such as videos~\cite{hodges2006sensecam, Dubourg2016SenseCamAN, Mann1996WearableTC}, audios~\cite{Hayes2004ThePA, Vemuri2004AnAP}, or bio-signals~\cite{chan2020biosignal}, to augment the memory database.
For example, recent work such as Memoro developed a wearable, audio-based device that continuously records users' conversations and enables memory suggestions in real-time, either through explicit queries or query-less contextual cues~\cite{chi2024memoro}.
Differently, \codename focuses on the ``post-capturing'' stage, utilizing already-existing memory data (e.g., photos and videos users have already captured). It addresses challenges in processing, annotating, and augmenting captured memories with contextual information.

Prior work in natural language processing (NLP), computer vision (CV), and information retrieval (IR) has studied methods of augmenting people's memory. 
Perhaps the most related is QA on egocentric videos, which are also a form of personal data.
Representative tasks include episodic memory retrieval~\cite{gurrin2014lifelogging, Grauman2021Ego4DAT, engel2023projectarianewtool}, where the system, given a long egocentric video and a query, localizes the answer within the video.
However, these datasets differ from the captured data targeted by \codename. The main challenge in egocentric videos is filtering through large, often noisy data, using data-driven approaches to train models for feature extraction. 
In contrast, captured memories represent a smaller, intentionally collected dataset, where the challenge lies in integrating scattered contextual information across multiple implicitly related memories.
Therefore, \codename employs a taxonomy-based method to augment existing data without the need for specific model training, improving QA performance.

\subsection{Multimodal Question Answering}

Over time, natural language QA research has shifted to more complex settings, including QA across different modalities (e.g., images~\cite{Agrawal2015VQAVQ, Goyal2016MakingTV}, videos~\cite{Yang2003VideoQAQA, Zhang2023VideoLLaMAAI, Maaz2023VideoChatGPTTD}, tables~\cite{Zhang2023MPMQAMQ} or knowledge graph~\cite{Yasunaga2021QAGNNRW, Huang2019KnowledgeGE}), QA on large datasets \cite{lewis2021retrievalaugmentedgenerationknowledgeintensivenlp, Chen2017ReadingWT} and tasks that require multi-hop reasoning~\cite{Yang2018HotpotQAAD, mavi2024multihopquestionanswering}.
Recent advancements in large language models (LLMs) and multimodal foundation models (e.g.,~\cite{liu2024llavanext, liu2023improvedllava, liu2023llava}) have enabled improved reasoning and answer generation over large, multimodal datasets.
This is similar to \codename's use case as answering personal questions requires handling large amounts of captured memories and performing complex reasoning.
Prior work has used retrieval-augmented generation (RAG) workflow~\cite{lewis2021retrievalaugmentedgenerationknowledgeintensivenlp}, which retrieves relevant information from external datasets based on a query and then generates output using the retrieved results. 
For example, MuRAG leverages RAG to answer open questions via retrieving related information from databases of images and text~\cite{Chen2022MuRAGMR}.
VideoAgent leverages structured memories processed from long videos to accomplish  video understanding tasks~\cite{fan2024videoagent}.
However, these methods rely on datasets already rich in context (e.g., Wikipedia\footnote{https://www.wikipedia.org/})
and improvements are often achieved by designing new query augmentation~\cite{chan2024rqraglearningrefinequeries} and retrieval workflows such as Self-RAG~\cite{asai2023selfraglearningretrievegenerate} and tree-based retrieval~\cite{sarthi2024raptorrecursiveabstractiveprocessing}.

\begin{colortext}
More recently, GraphRAG introduced a data augmentation approach that extracts a knowledge graph from raw data to tackle tasks requiring higher-level understanding, such as query-focused summarization~\cite{edge2024localglobalgraphrag}.
While we do not explicitly employ a graph data structure in \codename, we adopt GraphRAG's \textit{structured, hierarchical} approach for RAG-based tasks and extend it with taxonomy-based augmentation informed by insights from a diary study to enhance retrieval results on personal captured memories.
Finally, when it comes to QA system design, Jim Gray proposed the “20 queries” heuristic that optimizes for answering a core set of questions to address the long tail distribution of potential queries~\cite{szalay2008jim}.
We adopt the same design principle and replace the specific rules with our contextual information taxonomy.
\end{colortext}

\subsection{Applications Utilizing Contextual Information}

Contextual information has long been important in HCI research from early mixed-initiative systems~\cite{horvitz1999principles} to recent agentic workflows~\cite{jaber2024cooking}.
Over the past few years, there has been a surge in the usage of AI and LLMs in the HCI community to extract contextual information from processing raw multimodal information.
For example, Li \etal studied how visually impaired people cook and emphasized the importance of conveying contextual information to users through multimodal models~\cite{Li2024ACI}.
Additionally, Human I/O leverages egocentric perceptions of users and detect situational impairments through reasoning on the multimodal sensing data~\cite{Liu2024HumanIT}.
GazePointAR develops a context-aware voice assistant to disambiguate users' intent when interacting with real-world information~\cite{Lee2024GazePointARAC}.
OmniActions categorizes digital follow-up actions on real-world information and provides proactive action prediction based on perceived context~\cite{omniactions}.
These system utilized off-the-shelf multimodal models to process raw sensory data and leverage the reasoning capabilities of LLMs to infer the semantic context.
\codename builds on this approach by applying these AI techniques to extract and integrate semantic context scattered across various unstructured, raw captured memories.
This augmentation enhances users' memory databases, enabling them to answer personal questions about their memories through natural language queries.

\section{DIARY STUDY: UNDERSTANDING USER QUERIES}
While single captured memory often lacks essential contextual information, 
\codename proposes to augment such memories by extracting and inferring semantic context from other explicitly or implicitly related memories.
%
To understand how to effectively augment captured memories, we need to answer the following research question:
\vspace{-0.2mm}
\begin{itemize}[leftmargin=6.5mm]
    \item[\textbf{RQ}:] What contextual information is essential to integrate with captured memory instances to ensure accurate retrieval in response to user queries?
\end{itemize}
This question is important as ``context'' is a broad term, and thus the focus should be on categorizing and identifying the most effective contextual information that enables accurate and meaningful responses to the types of queries users generate when reflecting on past experiences.



\subsection{Method}
To answer the research question above, we conducted a diary study, a methodology that enables participants to log data whenever need arose \cite{diarystudy2008}. 
Specifically, we adopted the \textit{snippet-based technique} proposed by Brandt~\etal \cite{snippet10.1145/1240866.1240998}.
We asked participants to log queries on their past memories only when they had real intent under a genuine context,
rather than brainstorming potential questions they might ask to retrieve specific past memories. 
This approach enabled us to collect spontaneous, authentic queries that users have in real-world scenarios.

We collect the data including:
\one the queries participants would use to retrieve or ask about their past memories, 
\two the reasons and contexts of these queries (e.g., wanting to show a past experience while chatting with a friend) and
\three (optional) whether they were able to retrieve the corresponding memories from their album, and if so, how they did it (e.g., by scrolling through the photo album).







\subsection{Participants}

32 participants (14 male, 17 female, and one non-binary) were initially recruited through an online RSVP form distributed via the X platform\footnote{https://x.com/}. 
Participants came from North America and Asia. 
11 participants reported using Android devices, while the remainder used iOS devices in their daily life.
Additionally, 16 participants reported actively logging their daily lives, 13 regularly logged important events and memorable experiences, two logged only essential information, and one seldom logged their lives.
While participants were compensated based on their participation (\$50 for full participation), they were not required to log a specific number queries each day or over the entire study period.
This approach was intentional, as we did not want to require them to generating queries artificially.

\subsection{Data Summary}
\label{sec:data_summary}
During the diary study, one participant opted out during the first week, and two participants did not log any queries. Of the remaining participants, seven stopped logging after the first week. The rest remained active until the end of the study. As a result, we collected a total of 299 queries. On average, each participant contributed 10.27 queries ($\mathrm{SD} = 6.09$).
The highest number of queries from a single participant was 25
\begin{colortext}
and the lowest was 3.
\end{colortext}


From the collected queries, we identified three types of query: (1) direct content queries (75 queries), (2) context-based filters (28 queries), and (3) hybrid queries (191 queries).
The remaining five queries fell outside of these categories as participants attempted non-memory-related tasks like``Mark yesterday pictures as favorites''.


\paragraph{\textbf{Direct content queries}:}

These queries aim to get direct answers that can be retrieved by searching for memories via description (e.g., “\textit{skateboarding in a tie-dye shirt}”) or rely on information explicitly contained within a single captured instance (e.g., “\textit{What is my driver's license number?}”). 
This type of query \textbf{does not} require additional context not contained in a single captured memory.

\paragraph{\textbf{Contextual filters}:}
These queries focus on retrieving memories based on specific contexts, such as time, location, or event. 
For example, a query like “\textit{All the photos in Hawaii}” might only require filtering based on metadata like location. 
However, for more complex queries such as “\textit{All the photos from my graduation ceremony}”, 
it \textbf{does} require a deeper synthesis of multiple interconnected memories to reconstruct the context surrounding the event. 

\paragraph{\textbf{Hybrid queries}:}
These queries are more complex, combining both direct content queries and contextual filters. 
For example, a participant asked ``\textit{Which meat did I order the last time I came to this Japanese BBQ restaurant?}''
Answering such a query typically requires a \textbf{multi-hop} process: (1) filter all captured memories under the specific context (e.g., dining in this Japanese restaurant) and (2) analyze the filtered data to generate the final result.

\subsection{Analysis}
Inspired by the psychological memory theory \cite{tulving2002episodic}, 
our data summary indicates that 74.4\% of the queries (contextual filters + hybrid queries) require more than just querying the direct content. The complexity in these queries require integration of contextual information in captured memories for accurate processing and filtering. Therefore, we take a step further to build a taxonomy of contextual information in user queries to inform the design of \codename.

To identify this essential contextual information, two researchers on the team independently analyzed the logged queries.
They coded, filtered, and categorized the types of context required to filter captured memories and better answer the queries. Their results were compared, and discrepancies in categorizations, hierarchy, naming, and granularity were discussed and resolved.

\section{TAXONOMY OF CONTEXTUAL INFORMATION}

In this section, we present the taxonomy built from analyzing user queries. We identified three key types of contextual information that can be integrated with captured memories: (1) atomic context, (2) composite context, and (3) semantic knowledge.

\subsection{Atomic Context}

\begin{table*}[]
    \centering
    \caption{Categorization and examples of atomic and composite context }
    \renewcommand{\arraystretch}{0.5}
    \begin{tblr}{row{3-9}={5mm},row{11}={7mm},column{1}={22mm,c},column{2}={68mm,c},column{3}={67mm,c},colspec={|Q[c]|Q[c]|Q[l]|},rowspec={|Q[m]|Q[m]|Q[m]|Q[m]|Q[m]|Q[m]|Q[m]|Q[m]|Q[m]|Q[m]|Q[m]|}}
    \textbf{Category}   & \textbf{Definition}  & \textbf{Exemplar queries} \hl{\;\;\;} refers to contextual cues    \\
    \SetCell[c=3]{c,m} \textbf{Atomic context} \\
    \textit{Temporal info} & Specific time period or particular time of the day & {``\textit{What boba tea did I drink \hl{last week}?}'' \\ ``\textit{What is my routine \hl{in the morning}?}''}    \\
    \textit{Geographical info} &  Location data such as city names or venue details   &   ``\textit{How many churches did I visit \hl{in Barcelona}?}'' \\
    \textit{People} & Individuals present in the captured memories & ``\textit{Find the photo of \hl{me and my grandpa} last year.}'' \\
    \textit{Visual elements} & Other directly sensible elements, including animals, physical objects, or specific visual features &  
    {``\textit{My photo with \hl{short hair} last year.}'' \\ ``\textit{Photo of \hl{my dog} when he was a puppy.}''}\\
    \textit{Environment} & Inferred environment based on the content & ``\textit{\hl{Gym} selfies from last year.}'' \\
    \textit{Activities} & Actions or activities inferred from the content & ``\textit{How many \hl{cardio session} did I complete last month?}'' \\
    \textit{Emotion} & Subjective emotion or emotional cues & ``\textit{My \hl{happiest moment} last year}'' \\
    
    \SetCell[c=3]{c,m} \textbf{Composite context} \\
    - & Combination of multiple \textbf{atomic contexts} & ``\textit{Who did I ski with in \hl{the lab retreat} last year}'' \\
    \end{tblr}
    \label{tab:atomic_context}
\end{table*}

\begin{figure}
    \centering
    \includegraphics[width=0.8\linewidth]{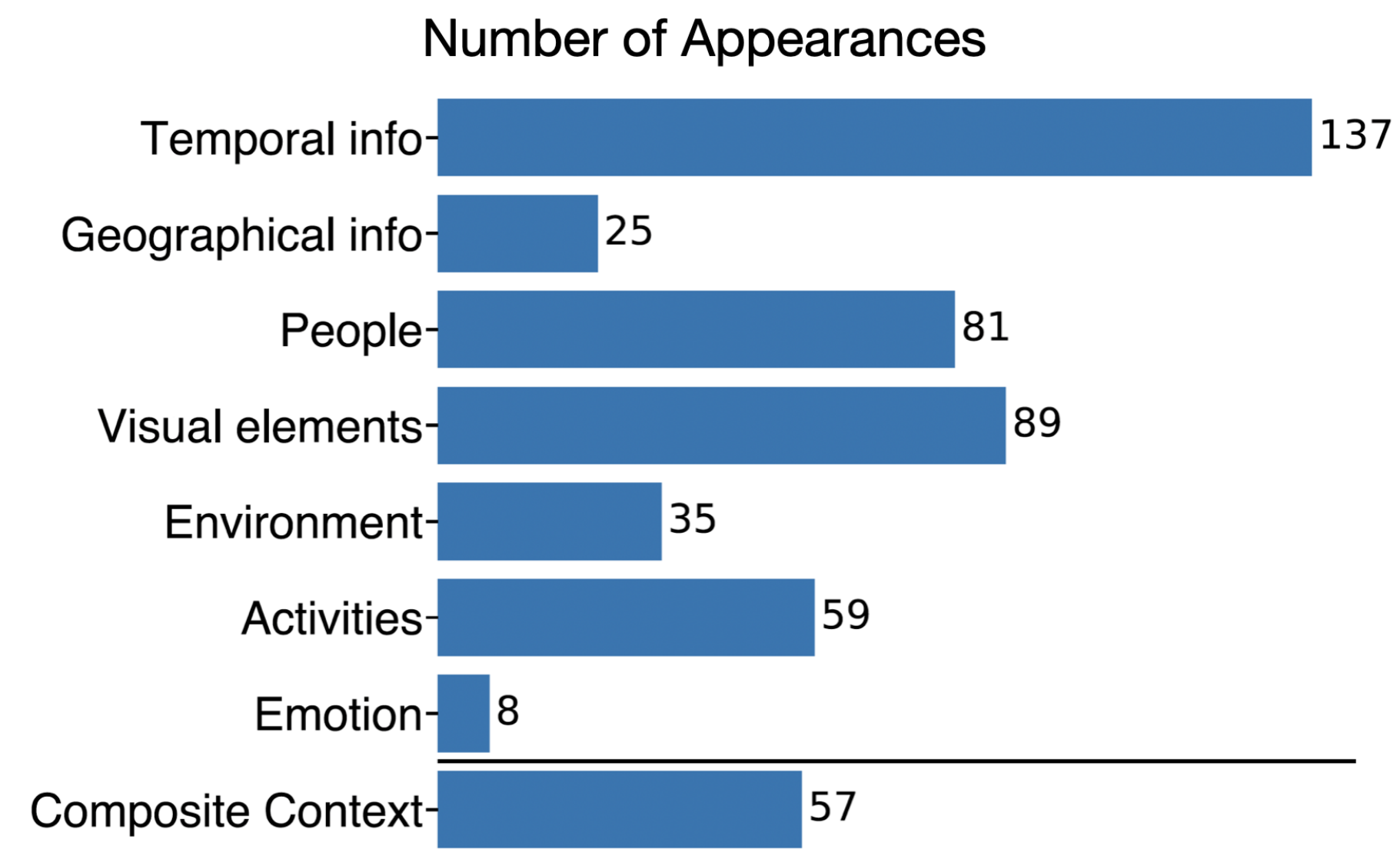}
    \caption{Number of appearances of each types of context (atomic and composite) in the logged queries. Note that a query may contain multiple types of categories, such as ``\textit{What \hl{boba tea} did I drink \hl{last week}?}''}
    \label{fig:context_count}
\end{figure}

Atomic context refers to contextual information typically obtainable from a single captured memory. 
This includes data directly from metadata, sensed from visual and auditory content, or inferred from the content itself.
Table~\ref{tab:atomic_context} shows the seven types of atomic contexts categorized from the queries. 
Among them, temporal information and geographical info can be directly obtained from the memory media's metadata.
People and visual elements typically require facial recognition or other vision models for detection. Environment, activity, and emotion are more implicit and require reasoning based on the content (e.g., a photo of a menu may suggest the person is in a restaurant).
The number of appearances of each category is shown in Figure~\ref{fig:context_count}.

\subsection{Composite Context}

Composite context is how people remember and refer to past experiences, such as "\textit{Who did I ski with in \textbf{the lab retreat} last year?}" 
These contexts can range from significant events like a wedding or a conference trip to smaller incidents like hanging out with a friend or a day trip to Seattle.
Specifically, composite context is defined as \textbf{a combination of multiple atomic contexts}.
For example, the composite context ``lab retreat'' encompasses atomic contexts including ``February, 2024'' (temporal), ``Lake Tahoe, California'' (geographical), and ``hanging out with labmates'' (activity). 


While atomic context is typically available within a single captured memory, composite context requires integrating multiple memories to understand the connection between them.
Since an individual's captured memories are linear on the timeline, memories related to a specific event tend to cluster closely together. 
We leveraged this \textbf{temporal proximity} to identify and extract various composite contexts from the raw captured memories. For a detailed discussion of this approach, please refer to Section \ref{sec:composite_context}.



\subsection{Semantic Knowledge}

In psychology theories, semantic knowledge refers to the general world knowledge that humans accumulate over time \cite{tulving2002episodic, mcrae201314}, distinct from episodic memories that are tied to specific experiences and events.
Similarly, we can generate semantic knowledge from a user’s captured memories, providing broader insights of the user's past experiences.
For example, patterns like “Jason has a habit of going to the gym 3-4 times a week” can be inferred from multiple captured memories.
Such patterns are helpful in answering queries that not necessarily require specific knowledge such as ``\textit{How often do I go to the gym in April?}''

\begin{figure*}
    \centering
    \includegraphics[width=1\linewidth]{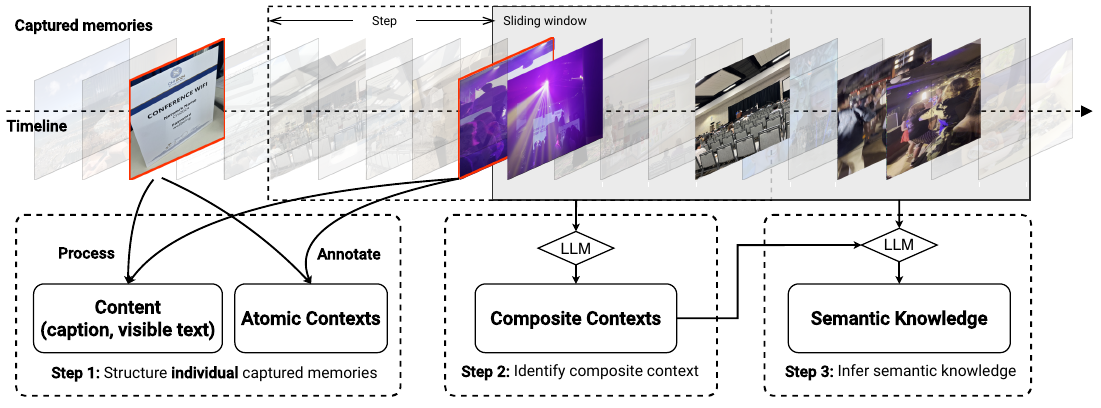}
    \caption{Augmenting captured memories involves three steps: (1) structuring memories by processing content and annotating with atomic contexts; (2) identifying composite context through sliding windows; (3) inferring semantic knowledge from the structured memories and identified contexts.}
    \label{fig:memory_anno}
\end{figure*}

\section{OMNIQUERY: AUGMENTING CAPTURED MEMORIES}








Informed by the generated taxonomy, \codename employs a \textbf{query-agnostic} preprocessing pipeline to augment existing captured memories.
The pipeline extracts scattered contextual information from interconnected captured memories, synthesizes it, and augment each memory with the enhanced context.
Specifically, the augmentation pipeline involves three steps (as shown in Figure~\ref{fig:memory_anno}):
(1) structuring individual captured memories via processing their content and annotating with atomic contexts,
(2) identifying and synthesizing composite contexts from multiple captured memories using sliding windows, and
(3) inferring semantic knowledge from multiple captured memories and the identified composite contexts. 

\subsection{Step 1: Structuring Individual Captured Memories}

\begin{figure*}[t]
    \centering
    \includegraphics[width=0.8\linewidth]{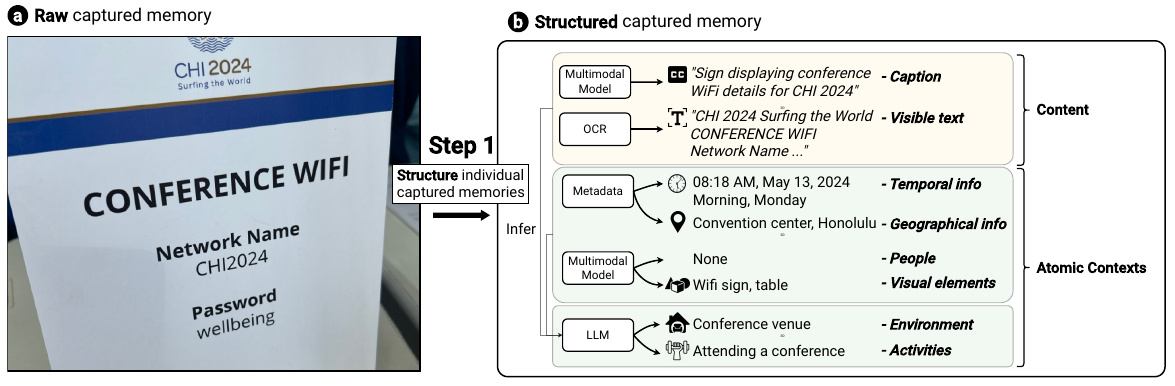}
    \caption{An example of structuring an individual captured memory (a photo of the Wi-Fi details of CHI 2024 conference).}
    \label{fig:step_one_example}
\end{figure*}

Raw captured memories are often unstructured and lack contextual annotation~\cite{sedlakova2023challenges}. 
In this step, \codename structures each captured memory, making it easier to analyze and extract information.
Figure~\ref{fig:step_one_example} shows an example of structuring an single captured memory, which involves two key parts:
(1) processing and understanding the content of the memory and (2) annotating the memory with atomic contexts.

\paragraph{\textbf{Processing content}}
Content of a captured memory includes an overall description of the memory as caption, visible text in the image, and transcribed speech (for videos, not shown in Figure~\ref{fig:step_one_example}).
Specifically, \codename leverages multimodal models to process and generate image captions, performs optical character recognition (OCR) to recognize visible texts, and uses audio-to-text models to transcribe speech.

\paragraph{\textbf{Annotating atomic contexts}}
With the content processed, \codename annotates each captured memory with each type of atomic context.
As shown in Figure~\ref{fig:step_one_example}b, \codename extracts the temporal and geographical information from the metadata and uses multimodal models to detect people and other visual elements.
Then \codename synthesizes the processed information and infers the environment and activities. 
For example, based on a photo of a sign displaying conference Wi-Fi details, \codename infers that the user is likely attending a conference (activity) and is at the conference venue (environment).
Note that due to the subjective nature of emotions that often requires user input, emotion inference is excluded from the current implementation.

\subsubsection{Indexing}

After each captured memory is structured, it is indexed and stored in a database.
Additionally, the annotations in textual format are encoded into text embeddings to enable vector-based search during the retrieval process.
In the database,
each data entry corresponds to a captured memory with both the original media (e.g., photo, video), its structured annotations in natural language, and text embeddings.




\subsection{Step 2: Identifying Composite Context}
\label{sec:composite_context}

\begin{figure*}
    \centering
    \includegraphics[width=0.95\linewidth]{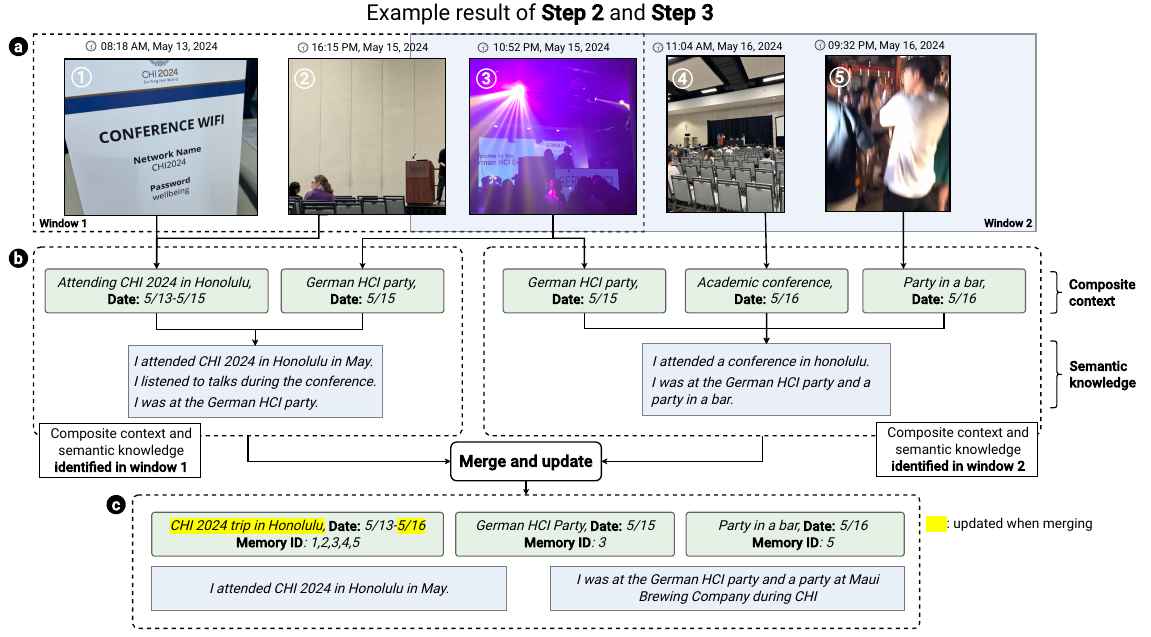}
    \caption{An example of using sliding windows to identify composite contexts and infer semantic knowledge: (a) two consecutive sliding windows; (b) composite contexts and semantic knowledge generated in each window; (c) merging results of the windows.}
    \label{fig:sliding_window}
\end{figure*}

As captured memories are recorded in a linear manner along a personal timeline, those interconnected through semantic contexts often cluster closely together.
For example, memories related to CHI 2024 are likely to occur during the event itself.
Taking advantage of this \textbf{temporal proximity}, \codename adopts a sliding window approach to analyze potentially interconnected memories scattered in segments for composite context identification.

As shown in Figure~\ref{fig:sliding_window}a, a static window size of seven days is used in our current implementation.
The inference is performed via an LLM, in which the input is the structured annotations of these memories and the output is the identified composite contexts along with their start and end dates and the associated captured memories (Figure~\ref{fig:sliding_window}b).
To account for cases where composite contexts are split in half, we use a step size (4 days in the current setup) smaller than the window size, allowing for overlap and comprehensive processing.
For longer composite contexts (e.g., lasting more than two weeks), each segment of the context is identified separately within the sliding windows and then merged into a single composite context. Additionally, any duplicated composite contexts caused by the overlap between sliding windows are also merged to avoid redundancy (Figure~\ref{fig:sliding_window}c).
\begin{colortext}
Note that we determined the window size heuristically. A longer sliding window can better capture extended events or patterns, while it may underperform on shorter contexts due to redundant information. 
One way to optimize is to make the size dynamic, adjusting it based on the density of activities in a given period. This approach would require a fixed dataset for experiments. We further discuss this in Section~\ref{sec:fixed_data}.
\end{colortext}

Specifically, as opposed to including detailed predefined categories (as with atomic contexts) in the prompt for LLMs, we adopt the few-shot prompting technique \cite{brown2020languagemodelsfewshotlearners}, providing examples of composite contexts summarized from the collected questions in the prompt.
For the detailed prompt, please refer to Appendix~\ref{app:prompt_context}.

\paragraph{Explicitly mentioned contexts.} Some composite contexts are \textbf{explicitly} mentioned in the captured memories. 
For example, a screenshot of a flyer may reference the upcoming "CHI 2024" event happening next month, or a transcribed conversation might discuss a "Hawaii trip" that took place the previous year.
We leverage \begin{colortext}LLMs' pretrained world knowledge\end{colortext} to differentiate between atomic contexts and composite contexts. 
For example, ``a workout session'' is identified as an activity (atomic context) because, based on world knowledge, it is more likely to refer to this activity alone.
In contrast, ``CHI 2024'' is recognized as a composite context, as it likely involves multiple interconnected atomic contexts.
Such identified composite contexts are either merged with an existing composite context (e.g., if "CHI 2024" has already been identified) or directly added as a new composite context if it is unique.

\subsection{Step 3: Inferring Semantic Knowledge}
Different from composite contexts, semantic knowledge focuses on high-level general knowledge rather than specific memory details. 
In the scenarios of personal memory, semantic knowledge refers to personalized knowledge distilled from an individual's past, as opposed to general knowledge (e.g., ``the capital of France is Paris.'').
For example, if a person's captured memories contain photos of attending CHI 2024, the distilled semantic knowledge might be, "The person attended the CHI conference in Honolulu in 2024."
Additionally, semantic knowledge goes beyond summarizing past events. 
It also encompasses inferred patterns and facts about the individual’s behavior or preferences. 
For example, a chat message mentioning Jason’s birthday could infer that "Jason's birthday is on [SPECIFIC DATE]." Similarly, analyzing multiple grocery shopping receipts that consistently include lactose-free milk could lead to the inference that the user is possibly lactose intolerant.
\begin{colortext}
We would like to note that, while inspired by human semantic knowledge as defined in psychology \cite{tulving2002episodic}, the semantic knowledge referenced here is slightly different. 
In \codename's setting, the semantic knowledge is objective and derived solely from the content of the memories.
In contrast, human semantic knowledge is more implicit, containing broader associations and longer-term effects that might go beyond just the observable data.

\end{colortext}


Semantic knowledge is inferred in each sliding window, while also taking into account the identified composite contexts (in Step 2) to gain higher-level understandings of the user's past and generalized information (Figure~\ref{fig:sliding_window}b bottom).
The output is a list of inferred declarative semantic knowledge independent from specific memories.
The instructions provided to the model are specifically tailored to guide the inference process toward overarching patterns and trends rather than specific event details.
The detailed prompt for identifying semantic knowledge can be also found in Appendix~\ref{app:prompt_knowledge}.
Each inferred entry of semantic knowledge is either merged with existing entries or added to the knowledge list if new.

\subsection{Implementation Details}
\label{sec:implementation}



To deduplicate images and videos as people tend to capture similar content multiple times, 
we use CLIP \cite{radford2021learningtransferablevisualmodels} to encode images (or the first frame of videos) into embeddings, calculate the similarities between images, and merge those with the similarity above 0.85. 
We use the Google Cloud Vision API\footnote{https://cloud.google.com/vision/docs/ocr} for OCR to detect text in images and OpenAI's Whisper model\footnote{https://github.com/openai/whisper} for audio-to-text conversion. 
Note that Whipser is known for hallucination when there is no speech in the audio, thus we applied further data cleaning to validate the transcribed result using OpenAI's \texttt{GPT-4o-mini}.
For other visual processing,
We use \texttt{GPT-4o} handles multimodal sensing, including identifying people and visual elements in images and generating scene descriptions.
For video processing, as a proof-of-concept, we consider only the first 10 seconds of each video, sampling 10 frames to be analyzed by \texttt{GPT-4o} for content understanding.
Text is encoded into embeddings using OpenAI's \texttt{text-embedding-3-small} model. Currently, we utilize a custom vector database and matrix-based similarity search implemented with NumPy in Python. However, for real-world applications, more advanced vector databases (e.g., Pinecone\footnote{https://www.pinecone.io/}) would be necessary to handle larger volumes of personal data.
\section{OMNIQUERY: QUESTION-ANSWERING SYSTEM}






With captured memories augmented with contextual information, \codename adopts a RAG architecture for the question answering system.
RAG-based systems are effective in handling large datasets and mitigating hallucination issues by retrieving relevant content and grounding the generated results in this retrieved information. This approach ensures that the output is both relevant and accurate, leveraging specific data rather than relying solely on the model's internal knowledge.
This approach is chosen because, on average, personal captured memories often exceed 30,000 photos and videos (as reported by participants in our diary study), which exceeds the limit of most foundation models nowadays.

As shown in Figure \ref{fig:qa_system}, 
given an input query, \codename first applies a \textit{taxonomy-based} augmentation of the query by disambiguating and decomposing it into specific contextual elements (Figure~\ref{fig:qa_system}a). 
Then, it retrieves the relevant captured memories from the structured captured memories and the composite contexts, along with related knowledge from the list of semantic knowledge (Figure~\ref{fig:qa_system}b). 
The retrieved memories and knowledge, along with the augmented query, are then sent to an LLM to generate a comprehensive answer (Figure~\ref{fig:qa_system}c).
We discuss detailed implementations of each step below.



\subsection{Taxonomy-Based Query Augmentation}


\begin{figure*}
    \centering
    \includegraphics[width=0.85\linewidth]{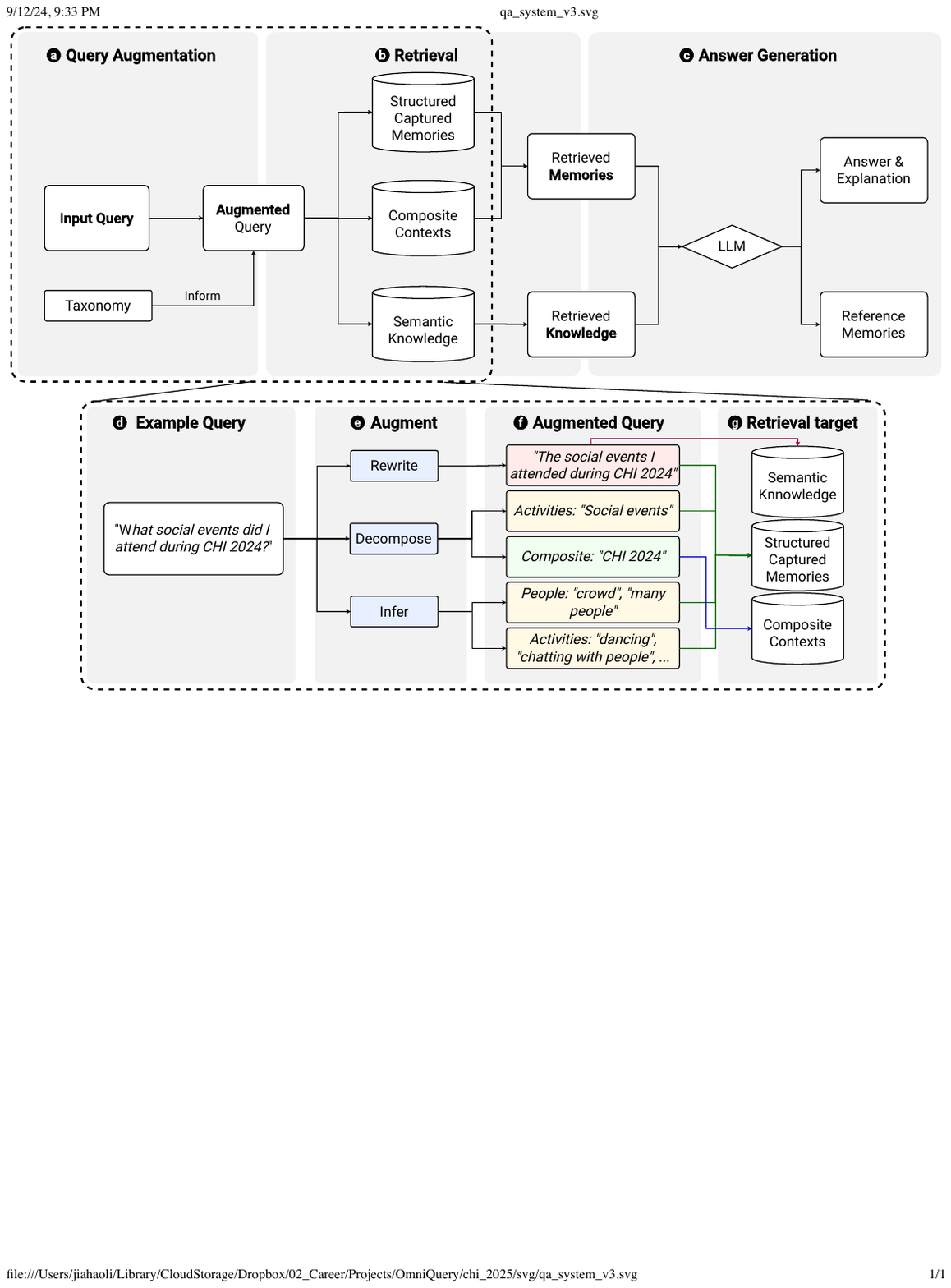}
    \caption{
    The question-answering system consists of: (a) taxonomy-based query augmentation by decomposing and inferring contextual information; (b) retrieving memories and semantic knowledge; (c) generating answers with an LLM using referenced memories. Specifically, given an input query (d), \codename first augmens it via rewriting and decomposing and inferring contextual information~(e). The augmented query with different types of decomposed contextual information (f) is used to retrieve captured memories from different memory and knowledge storage (g).}
    \label{fig:qa_system}
\end{figure*}

As mentioned in Section~\ref{sec:data_summary}, most user queries tend to be hybrid in nature or require contextual information. 
This means that directly searching based solely on the content of captured memories often results in an incomplete or insufficient retrieval of relevant memories.
To enhance the retrieval process, \codename adopts the query refinement approach~\cite{chan2024rqraglearningrefinequeries} to augment the queries. 
This query augmentation process is also informed by the taxonomy of contextual information and it involves 
\begin{enumerate}[leftmargin=4.5mm]
    \item \textbf{Rewriting the query} to declarative format to improve search accuracy of vector-based similarity matching;
    \item \textbf{Decomposing the query} to extract necessary contextual filters, such as time, location, or events, which are grounded in the taxonomy. Note that only explicitly mentioned temporal contexts like ``... last week'' will be recognized temporal filters. Phrases like ``... during CHI 2024'' are part of a composite context and thus not counted as a temporal filter;
    \item \textbf{Inferring potential related contexts} that may not be explicitly mentioned in the query but can enhance the filtering process also grounded in the taxonomy.
\end{enumerate}

\noindent
For example, as shown in Figure~\ref{fig:qa_system}d-g, the query ``\textit{What social events did I attend during CHI 2024?}'' is rewritten into a declarative format of ``\textit{The social events I attended during CHI 2024}''. 
\begin{colortext}
We leveraged an LLM to classify the mentioned contexts in the query as either atomic or composite. 
Detailed prompts are provided in Appendix~\ref{app:prompt_query_aug}.
\end{colortext}
Since ``CHI 2024'' is explicitly mentioned and identified as a composite context, it is extracted and labeled with the appropriate composite context tag. ``Social events'' is also extracted and identified as an atomic context (activities).
Additionally, because ``social events'' might include various activities like parties, dancing, or casual conversations and involve multiple people, \codename infers the relevant atomic contexts (people and activities) and annotates them in the corresponding context category.

\subsection{Retrieving Relevant Augmented Memories}




The decomposed augmented query is used to comprehensively retrieve relevant augmented captured memories.
The augmented captured memories consist of the structured captured memories (with processed content and annotated atomic contexts), the list of identified composite contexts, and the list of semantic knowledge.
\codename uses the decomposed components from the augmented query to perform a multi-source retrieval, pulling related memories from each of these sources. The results are then consolidated into a comprehensive list of relevant memories, which are used to generate an accurate and detailed answer for the user's query.\\

\noindent
\textit{\textbf{Declarative query $\rightarrow$ Semantic knowledge \& processed content.}}
The declarative query is encoded into text embeddings to search for both the semantic memories and processed content (caption and visible text) of the captured memories.
This initial search step focuses on finding knowledge and memories directly related to the input query, without incorporating additional contextual filters.\\

\noindent
\textit{\textbf{Decomposed atomic contexts $\rightarrow$ Annotated atomic contexts.}}
Each element of the decomposed atomic contexts (both extracted or inferred) is encoded into text embeddings and searched through the corresponding categories in the structured captured memory database.
For example, if the query involves activities like "party" and "dancing," \codename searches for captured memories annotated with similar activities. Any memories that have been annotated with related or similar activities will be retrieved, ensuring that relevant memories are included in the results.
Additionally, \textit{temporal contexts} apply a \textbf{strict} filter, excluding memories outside the specified time frame (e.g., ``last month'') from the retrieval process.\\

\noindent
\textit{\textbf{Decomposed composite contexts $\rightarrow$ Identified composite contexts.}}
Any composite context decomposed in the augmenting process is also encoded into text embeddings and searched through the list of identified composite contexts. 
All captured memories linked to the semantically similar composite contexts are retrieved. This ensures all memories related to the composite contexts are included. 
Additionally, \codename leverages an LLM to assess whether a composite context includes temporal constraints. For example, ``\textit{... during CHI 2024}'' implies a strict temporal filter, while ``\textit{photos related to CHI 2024}'' does not.

\subsection{Answer Generation}



The retrieved results is then sent to an LLM to generate the final answer.
Specifically, the input for the LLM consists of:
(1) the augmented query, 
(2) the retrieved semantic knowledge from the list,
(3) all the retrieved captured memory entries from the annotated database, including both the memory content and its associated contextual annotations.


The model analyzes and reasons which captured memories serve as references for the generated answer. 
These reference memories are also included in the final answer presented to the user. 
To enhance the reasoning process, \codename leverages chain-of-thought prompting~\cite{wei2023chainofthoughtpromptingelicitsreasoning}, ensuring the generation is more accurate and contextually rich (specific prompts in Appendix~\ref{app:prompt_answer}).


\begin{figure*}
    \centering
    \includegraphics[width=0.95\linewidth]{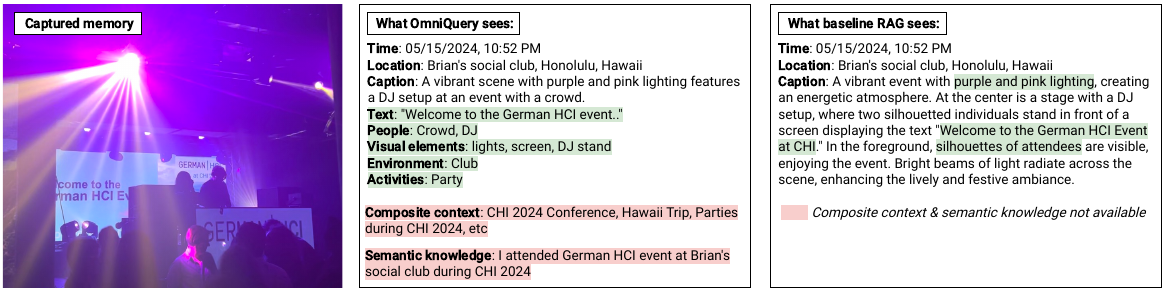}
    \caption{\begin{colortext}For each captured memory, the baseline RAG system lacks the contextual information extracted through the taxonomy-based augmentation used in \codename.\end{colortext}}
    \label{fig:baseline_comparison}
\end{figure*}

\section{USER EVALUATION}
We conducted a user evaluation to test \codename's capabilities in handling real-world personal data by comparing it against a baseline system implemented with a typical RAG structure for question answering. 
Both systems were deployed on the participants' local machine to protect their personal data.
In this section, we discuss the detailed evaluation process, metrics, and results, including quantitative results of the two systems' performances, representative examples, and qualitative feedback.

\subsection{Participants}

We recruited 10 participants, including seven from our diary study and three additional participants via word-of-mouth. 
They consented to the whole process, including that their filtered personal data will be processed via an API service. All 10 participants four male, six female, age range 22 - 29, $\Bar{x} = 25.3$, $\mathrm{SD} = 2.63$) were fluent or native English speakers. 
The participants rated their frequency of logging their daily lives as 
`Only record essential information' (1), `Regularly log important events and memorable experiences' (5) and `Actively log my daily life' (4). 
Each participant was compensated with \$50 for completing this study.

\subsection{Apparatus}

Two different systems were implemented in the user evaluation: the \codename pipeline and a baseline system for comparison. 
\begin{colortext}
The baseline RAG is designed to differ only in \codename's core contributions.
As shown in Figure~\ref{fig:baseline_comparison}, while the baseline RAG includes basic contextual information for each captured memory, such as time, location, and a detailed description of the scene, the key differences are:
(1) The contextual information remains “raw” and is not structured using the taxonomy derived in \codename, and
(2) it lacks composite context and semantic knowledge, as it does not extract contextual information from multiple related memories.
Additionally, in the question-answering phase, the baseline does not leverage taxonomy-based retrieval.
All other components, such as the base LLM and prompt structure, remain the same.
This design ensures the delta between \codename and the baseline highlights and evaluates \codename's core contributions: taxonomy-based augmentation and retrieval.
\end{colortext}
For the detailed implementation of the baseline, please refer to Appendix~\ref{app:baseline}.

\begin{figure}
    \centering
    \includegraphics[width=0.9\linewidth]{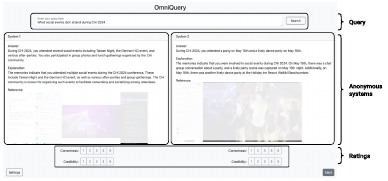}
    \caption{User interface used in the user evaluation.}
    \label{fig:user_interface}
\end{figure}

As shown in Figure~\ref{fig:user_interface}, in the studies, participants were presented with a single text input box similar to search engine input boxes. After they typed in the question, they would see two answers generated by the two systems in a randomized order. Two rating questions on the answer's accuracy and completeness were then shown under each answer with a scale of 1-5.

\subsection{Procedure}
The user evaluation has three stages: (1) system setup 
\begin{colortext}
on participants' local machines,
\end{colortext}
(2)
\begin{colortext}
personal 
\end{colortext}
data preparation, and (3) the main testing session.

\paragraph{\textbf{System setup}}
The participants were given the source code for \codename to install the back-end and a web application on their local machine.
They had the option of an online walkthrough session with the experimenters or following the setup instructions on a self-guided manner.

\paragraph{\textbf{Personal data preparation}}

\begin{table*}
\centering
\caption{Quantitative Results of \codename and Baseline, including UPA, UPC, and Accuracy (\%)
}
\renewcommand{\arraystretch}{1.05} 
\begin{tabular}{C{4cm}*{6}{C{0.8cm}}}
\hline

 &  \multicolumn{3}{c}{\textbf{\codename}} & \multicolumn{3}{c}{Baseline} \\ 
\cline{2-7}

Metrics & UPA  & UPC &  ACC  & UPA & UPC & ACC   \\ \hline

Direct content query (24)         &     4.42       &    4.13   & 83.3  & 3.67  & 3.46  & 62.5      \\ 
Contextual filter + Hybrid  (113)         &   3.89     &         3.85 & 69.0 & 2.93 & 2.83 &  38.9      \\ 


\hline
\begin{colortext}Queries from diary study (28)\end{colortext}        &   3.93     &     4.14 & 67.9 & 2.39 & 2.61 &  25.0      \\ 
\hline

\textbf{All} (137)     &      \textbf{3.98}        &     \textbf{3.90}  & \textbf{71.5}  & 3.06  & 2.94  & 43.1       \\
 
\bottomrule

\multicolumn{5}{l}{\footnotesize *ACC refers to accuracy, which is considered accurate when UPA $\geq$ 4 (mostly correct).} 
\end{tabular}
\label{tab:target_accuracy}
\end{table*}

\begin{table*}
\centering
\caption{Direct comparison between \codename and baseline 
}
\renewcommand{\arraystretch}{1.05} 
\begin{tabular}{C{4cm}*{4}{C{1.8cm}}}
\hline

 &  \multicolumn{4}{c}{\textbf{Comparison Win Rate (\%)}} \\ 
\cline{2-5}

Winner & \textbf{\codename}  & Baseline &  Tie  & Both are bad   \\ \hline

Direct content query (24)          &      50.0     &    8.3 & 33.3  & 8.3      \\ 
Contextual filter + Hybrid (113)         &    53.1    &    11.5  & 19.5  & 15.9           \\ 

\hline
\begin{colortext}Queries from diary study (28)\end{colortext}         &    60.1    &    7.1  & 14.3  & 17.9           \\ 

\hline

\textbf{All} (137)     &      \textbf{52.6}        &     10.9  & 21.9  & 14.6       \\
 
\bottomrule
\end{tabular}
\label{tab:win_rate}
\end{table*}
\begin{colortext}
To test \codename on local machines, the participants were asked to transfer a set of captured memories (both photos and videos) from their smartphones' albums to their laptops.
To further protect participants' privacy, 
\end{colortext}
they were instructed to manually review and filter out any content deemed sensitive or preferred to be excluded from the study.
\begin{colortext}
This process needs to meet two key requirements:
(1) the manual filtering should not become an excessive burden for participants, and
(2) the transferred data should be sufficiently large to simulate real-world usage, containing diverse contexts of distant memories.
It is impractical to include all participants' memories, which average 13630 files as collected from the diary study.
Thus we perform the following calculation to balance the two requirements:
\begin{equation*}
T \geq \frac{C}{K} \times \frac{1 + R}{1 + F \times R}
\end{equation*}

Where $T$ is the total number of files, $C$ the context limit, $K$ the token cost per frame, $R$ the relevant frame ratio, and $F$ the average frames per video. The equation balances manual filtering and data diversity.
The rationale behind the equation is that processing all memories at each query should exceed the limit of existing powerful models, and thus motivating the need for accurate retrieval to answer queries.
Specifically, the model \codename uses has a context window of \texttt{128K}\footnote{\url{https://platform.openai.com/docs/models\#gpt-4o}}, and the minimum token cost for processing each frame when employing low-fidelity understanding is \texttt{85}\footnote{https://platform.openai.com/docs/guides/vision}.
\textit{Ratio} represents the video-to-image ratio, which averages 0.12 ($\mathrm{SD} = 0.11$) based on our diary study survey.
As discussed in Sec.~\ref{sec:implementation}, the number of frames sampled per video during processing is 10.
This calculation results in a total number of files of \textasciitilde767. 
To expand the memory coverage within the participants' manual filtering capacity, we round this up to 1,000 files.





\end{colortext}

Depending on how frequently participants logged their daily lives, the \begin{colortext}memory coverage of the 1,000
\end{colortext}selected photos and videos spanned from \textit{one to four months}
\begin{colortext}
($\Bar{x} = 2.3$, $\mathrm{SD} = 1.03$).
While this coverage is smaller than what a smartphone album can cover in real-world scenarios (several years), it still represents a moderately distant memory range that is not too recent to demonstrate the capabilities of \codename.
\end{colortext}
These captured memories were then fed into \codename for the taxonomy-based query-agnostic augmentation process. 
The safety of the process data is ensured following the API's privacy protocol\footnote{https://openai.com/policies/privacy-policy/}.

\paragraph{\textbf{Main session}}



The main session lasts 45 minutes, in which the participants tested \codename using two types of questions: (1) questions logged during the diary study and (2) questions they generated during the session.
The participants checked on the diary study questions manually to determine if they could be answered using the filtered set of data on captured memories. Additionally, the participants were encouraged to brainstorm and use new questions that were potentially answerable using the filtered data to comprehensively test \codename.
\begin{colortext}
Note that participants have no access to the contextual information augmented by \codename. Instead, they follow a simple mental workflow: attempting to recall, asking the system, and verifying the answer.
\end{colortext}


In the question-answering procedure, \codename-generated answers are accompanied with answers generated from the baseline system implemented using RAG. 
Each system generated answers anonymously, and the participants compared and rated the results for both systems.
The answers and user ratings were recorded for quantitative analysis.
Throughout the process, the participants were asked to think aloud~\cite{nielsen1994usability}, and a brief interview was conducted at the end of the session to gather feedback and suggestions.
These results were recorded for qualitative analysis.

\subsection{Comparison Metrics}

After two answers were presented for a question, the participants were asked to rate the two answers.
We used the Chatbot Arena evaluation method~\cite{chiang2024chatbotarenaopenplatform}, where each participant compared answers from the two systems and selected the better one or marked it as a tie.
More specifically, for each question, each participant rated the \textbf{user perceived accuracy} (UPA) and \textbf{user perceived completeness} (UPC) of the answers from both systems.


The UPA score was rated on a scale from 1 to 5:
\textbf{1}: Completely wrong or invalid result;
\textbf{2}: Incorrect, but provides at least some insight that helps answer or further refine the question;
\textbf{3}: Partially correct, or contains a subset of correct answers (e.g., only listing one meal when asked about all meals eaten last week);
\textbf{4}: Mostly correct, but missing some minor details (e.g., missing one subway trip when asked how many times I rode the subway);
and \textbf{5}: Completely correct.
%
The UPC score focused on the completeness and credibility of answers, essential for assessing reliability. 
Many participant queries were complex, requiring explanations supported by captured memory data. 
For example, when asked ``How many meals did I have during my last New York trip?'' a system might get the number right but miscount specific meals. Participants reviewed filtered data and rated answers based on their perceived completeness and credibility.

\begin{figure*}[ht]
    \centering
    \includegraphics[width=1\linewidth]{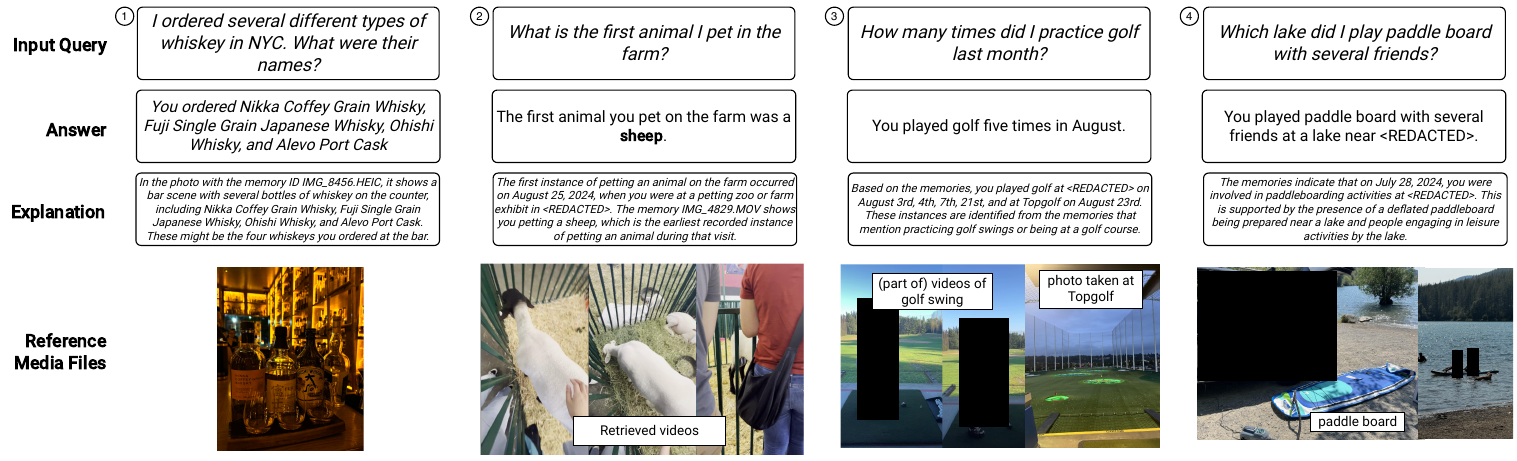}
    \caption{\begin{colortext}Four representative examples of using \codename to answer hybrid personal memory-related questions.\end{colortext}
    }
    \label{fig:result}
\end{figure*}


We also directly compare \codename and the baseline to analyze performance. 
If both have a UPA of 2 or lower, the result is labeled ``both are bad''. 
If at least one scores $\geq$3, the system with the higher UPA wins. In case of a tie on UPA, the system with the higher UPC wins; otherwise, it remains a tie.

\begin{colortext}

\section{Results}
We report the results of our user evaluation by presenting quantitative results of comparing \codename with the baseline, representative examples from the study, and insights gathered from the think-aloud protocol and exit interviews. 

\subsection{Quantitative Result}

\end{colortext}

The participants tested 137 queries in total during the main session. 
Among them, 28 were previously logged during the diary study. 
We manually labeled each tested query using the categorization and definition mentioned in section~\ref{sec:data_summary}.
As a result, 24 were categorized as \textit{direct content query} while 17 were \textit{contextual filters} and 96 were \textit{hybrid queries}.
We analyzed the performance metrics of both systems (\codename and baseline) using the scores rated by the participants.
Table \ref{tab:target_accuracy} and \ref{tab:win_rate} summarize our results.
In addition to presenting the average UPA and UPC scores, we calculated binary accuracy to evaluate whether the systems provided mostly correct answers. 
\begin{texthighlight}
An answer was considered accurate if its UPA score was equal to or greater than 4 (mostly correct) (Table \ref{tab:target_accuracy}).
\end{texthighlight}
We also present the ``comparison result" in Table~\ref{tab:win_rate}, which compares the two systems head-to-head on answering personal questions.

The result shows that, overall, \codename outperforms the baseline system in both the accuracy and completeness. 
Specifically, \codename achieves an accuracy of 71.5\%, outperforming the baseline by 28.4\%, winning the comparison 52.6\% of the time, and tying 21.9\% of the time.
For 14.6\% of the time, both results are bad.
We also present the results for different categories of queries.
The results indicate that simpler techniques like the baseline handle direct content queries reasonably well (62.5
\% accuracy, and winning or tying 41.6\% of the time).
While the baseline struggles with more complex queries such as contextual filters or hybrid queries (38.9\% accuracy, winning or tying 31.0\% of the time), \codename demonstrates it capabilities in effectively handling such queries (69.0\% accuracy, winning or tying 72.6\% of the time).
Specifically, for the queries logged during the diary study, \codename achieved results similar to its overall performance (67.9\% accuracy, and winning or typing 74.4\% of the time).



\subsection{Representative Examples}
We selected four representative examples tested by the participants in the evaluation, which are illustrated in Figure~\ref{fig:result}.

\begin{colortext}
\vspace{-1mm}
\begin{enumerate}[leftmargin=4.5mm]
  \item P3 wanted to recall the name of the whiskey they tasted during their trip, as they enjoyed it and wanted to check the price. \codename successfully retrieved the target memory (a photo of the four bottles) and generated a specific answer about the bottles they might have ordered.
  
  \item P6 wanted to organize the footage from their visit to a farm and asked about the first animal they petted there. \codename accurately retrieved the memories related to the composite context (``visit to the far'') and used the temporal order to generate the correct answer: the first animal petted was a sheep.
  \item P1 wanted to estimate how many times they had practiced golf in the past month to track their progress. \codename successfully retrieved all relevant memories, including videos of golf swings at the driving range and photos taken at Topgolf, and accurately generated the answer.
  \item P9 wanted to recall the name of the lake where they went paddling with friends. While no direct memory of paddling is captured, there are several related photos available, including a paddleboard being pumped next to a lake.  \codename successfully retrieved these memories and generated the answer using the metadata associated with them.
\end{enumerate}

\end{colortext}

\subsection{Qualitative Feedback and Findings}



All participants had experience using smartphone album search features, primarily for retrieving specific information like driver’s licenses or events such as trips, aligning with direct content queries and contextual filters (Section \ref{sec:data_summary}). However, they noted that existing tools are limited to finding specific objects and cannot handle more complex queries. Plus, some of our participants also anticipated for this to happen because they ``know what can be searched and what cannot be searched'' from these existing album search tools (P2).



In the studies, a lot more challenging questions were asked. For \textit{direct content queries}, it would be challenging to answer when the object is ambiguous or when the users can only describe the object and do not know its exact name. For \textit{contextual filter} and \textit{hybrid} or even more open-ended and subjective questions, existing searching tools are not comparable to \codename and the baseline at all because tools like iOS album search only return specific photos and videos without contextual understanding or filtering.

Here we further summarize the cases that are hard to be accomplished by existing tools. In comparison to the high-level question types provided in Section \ref{sec:data_summary}, we dive deeper into what these questions in the study were about and provide detailed examples.


\begin{itemize}[leftmargin=3mm]
    \item \textbf{Exploratory Search}: When users know some characteristics of what they are searching for but cannot specify the exact object. For instance, P1 asked, ``What churches did I visit in Barcelona?''
    
    \item \textbf{Look up and Locate}: When users know specific references or attributes about an item, such as date, location, or a person in the photo, and want to quickly locate the relevant media, such as ``Can you find the photo of me on a flyer on Instagram? (P4)''

    \item \textbf{Summarization Tasks}: Participants often need answers that summarize their collection of media, rather than finding a single item. For instance, P7 queried, ``Which subway stations in New York have art installations?''.

    \item \textbf{Comparative Questions}: Users sometimes want to compare different sets of media. 
    For example, P10 asked, ``Am I enjoying beach time more or hiking more?''

    \item \textbf{Open-ended and Subjective Questions}: Participants also asked questions that require interpretation or subjective judgment, which were even more challenging for existing tools. 
    For example, P5 asked, ``Given the photos I took, could you analyze what kind of person I am?''
    
\end{itemize}
\noindent
In the meantime, we want to emphasize again that the comparison between \codename and existing tools is conceptual, given that they serve different purposes and are designed differently in retrieving objects or answering questions. We provide this conceptual comparison to demonstrate the variety of questions \codename can support answering.

\subsection{Failure Cases}
Among the 137 queries tested, 25 failed due to ambiguity, missing contextual cues, information loss, retrieval redundancy, subjectivity, or unavailable memories. 
Please refer to Appendix \ref{app:failure_case} for detailed discussions.

\section{DISCUSSION}
In this section, we draw on implications from our studies to both discuss limitations and propose future work.
\begin{colortext}

\subsection{Curating Fixed Dataset for Benchmarking}
\label{sec:fixed_data}
Benchmarking is a widely used approach to evaluate whether new systems and algorithms achieve state-of-the-art performance on specific tasks.
It also enables ablation studies to assess the effectiveness of individual components of the proposed design.
Currently, \codename is evaluated as an integrated system to compare against another system to demonstrate the effectiveness of its overall design. 
As a future direction, curating a fixed benchmark dataset would allow for more granular evaluation of \codename's performance on personal question answering over multimodal captured memories.
This would enable deeper insights into how different design choices (both high-level and low-level) affect task performance through objective ratings. 
Additionally, it could enable testing multiple system parameters (e.g., sliding window sizes, top-K values for retrieval, or prompt designs for inferring contexts) to achieve the optimal performance. 
Furthermore, it would also allow ablation studies to assess the impact of individual components (e.g.,~query augmentation). We further discuss significant challenges of curating a fixed benchmark dataset for this personal data task in Appendix \ref{app:fixed_dataset}.

\end{colortext}

\subsection{From Chat Interface to Multimodal Interactions}
\label{sec:discussion_uncertainty}

As a system designed to answer user queries on their personal captured memory, \codename is currently designed in an ask-and-react manner to evaluate its efficacy in a lab-study setting. In our studies, the participants were excited about what \codename was capable of and gave feedback on having more multimodal interactions rather than just a chat interface. We recognize the potential of a more interactive \codename in the following ways:

\textit{\textbf{Multimodal Input and Output.}}
\codename could support multimodal inputs, including audio, images, and videos, to address limitations of text-based search. Many challenging cases for existing album search tools could benefit from this, such as locating an oddly shaped cup or matching dresses by color. Beyond retrieval, \codename could help users relive memories by visualizing captured data, enabling interactive exploration and annotation edits. This brings us closer to a “mind palace” style AI assistant.

\textit{\textbf{Error correction.}} 
In our studies, we observed the importance of enabling users to review and refine identified composite contexts and semantic knowledge. Participants expressed the need to correct errors when the system retrieved irrelevant information. For example, P9 asked about a K-pop store, but the system mistakenly included an Instagram screenshot of a Korean TV show. To address this, we propose integrating error correction mechanisms with explanatory insights, confidence levels, and a verification loop, allowing users to mark errors, refine results, and enhance system accuracy over time.


\textit{\textbf{Follow-up queries.}}
A key theme in our study is participants' need to refine queries or ask follow-up questions, with six out of ten mentioning the desire to clarify responses or narrow their searches iteratively. This was particularly relevant when errors were perceived, as discussed in Error Correction. To address this, we propose augmenting follow-up interactions with explanations and confidence levels to highlight uncertainties. A top-K retrieval strategy could also provide ranked answers for ambiguous queries, enabling iterative refinement. Future work could evaluate these approaches through a longitudinal study.

\subsection{Enriching Memory Data and Visual Intelligence}

At present, \codename primarily processes media from a smartphone's photo album as its main source for captured memories. However, these media alone provide a limited view of a user's broader personal knowledge. For example, in one of the study's failure cases, \codename struggled to infer personal relationships from social interactions captured in group photos. To enhance memory augmentation and improve retrieval accuracy, expanding \codename's data sources and visual intelligence is essential.

\begin{colortext}
\textit{\textbf{Integrating additional data structure and sources.}}
\end{colortext}
Personal knowledge extends beyond photo albums and exists across various applications. While our participants' photo albums included screenshots of emails, calendar events, and chat histories, these represent only a fraction of the broader personal information available in other communication and social interaction apps. 
Incorporating data from such sources could significantly enhance \codename's contextual understanding, allowing for more complex queries and richer memory retrieval.
\begin{colortext}
    While \codename already extracts hierarchically structured information from raw photo album data in the form of atomic and composite contexts, future work can explore using other explicit data structures such as graphs and trees to organize data from multiple sources.
\end{colortext}
Integrating these additional data sources also presents substantial privacy and ethical challenges. While our evaluations were conducted entirely on users' local machines and did not explore privacy-preserving implementations in detail, existing research efforts, such as those focused on differential privacy and on-device machine learning, offer promising directions for secure and privacy-aware deployment. Additionally, commercial tools like Apple Intelligence's private cloud computing serve as examples of ongoing progress in protecting user data while enabling advanced memory retrieval.

\textit{\textbf{Enhancing visual intelligence.}}
Queries related to social interactions remain challenging due to the current lack of advanced features like facial recognition for person identification. Future iterations of \codename could integrate such capabilities (with appropriate user consent), enabling the system to track individuals across various memories. This enhancement would support new use cases, such as monitoring social patterns or tracking progress over time, significantly improving the system's capacity for memory augmentation and retrieval. Additionally, we propose exploring the design and implementation of a comprehensive taxonomy of personal knowledge domains. This would allow users to selectively activate specific domains, such as enabling ``Social Interactions and Relationships'' to infer personal connections while disabling ``Personally Identifiable Information'' to prevent the system from processing sensitive data like IDs or SSNs in photos. This modular approach could enhance both user control and privacy.

\textit{\textbf{Augmenting with future AR technologies.}}
A limitation of personal memory capture is the potential for missed moments when users either forget or are unable to document an experience. As AR technology advances, \codename's memory augmentation and retrieval capabilities could be seamlessly integrated into AR systems, allowing for more passive and context-aware memory capture. AR devices could leverage real-time contextual triggers \cite{omniactions} to proactively surface relevant memories or information, offering proactive assistance in pervasive AR environments. This integration would enhance the user experience by making memory retrieval more intuitive and contextually relevant. However, such passive data capture raises even more significant privacy concerns, which will require future research into secure, privacy-preserving implementations to ensure the responsible use of AI in these settings.

\subsection{Preserving Privacy}
\label{sec:privacy}



As discussed above, protecting users' privacy is crucial in developing future personal AI assistants, including but not limited to handling personal data such as media in albums and chat and browsing histories. Users have limited control over how their data is handled and must rely on service providers' adherence to privacy protocols. In this subsection, we take a step further to discuss more robust and rigorous measures that should be adopted in real-world settings, where the immense amount of personal data makes approaches like manual filtering in \codename's evaluation infeasible.


One way is to incorporate more advanced data protection techniques, such as data anonymization~\cite{Majeed2021AnonymizationTF} and encryption~\cite{Nadeem2005APC}, while preserving the computational capabilities of large models via online computing.
The other approach is leveraging on-device computing, where all data processing occurs locally on the user's device, ensuring full control over users' own data.
Recent advances in model compression~\cite{Hohman2023ModelCI} have made it possible to run large model on smaller devices like smartphones.
As \codename is designed to be model-agnostic, it is able to work with different model sizes.
While smaller, compressed on-device model may result in reduced performance, future work should focus on developing curated datasets and benchmarks to rigorously evaluate \codename’s performance across different model sizes (e.g., LLaMAs~\cite{touvron2023llamaopenefficientfoundation} and Phi-3~\cite{abdin2024phi3technicalreporthighly}). This would provide a deeper understanding of how model size impacts privacy and system effectiveness.

\section{CONCLUSION}



We present \codename, a pipeline that enhances personal question answering on captured multimodal memories. Informed by an one-month diary study, \codename's design responds to real-world user queries and synthesizes a contextual taxonomy of captured memories. Our pipeline design of structuring individual captured memories and identifying composite context and semantic knowledge using a sliding window technique was used to develop a question-answering system, which outperformed a baseline RAG system in both perceived accuracy and completeness. Unlike existing research and commercial tools focused on intelligent image retrieval in smartphone albums, \codename is the first to tackle complex and nuanced personal queries, moving beyond simple object or information piece retrieval. With further attention to privacy-preserving measures, we believe \codename holds significant potential to evolve into a comprehensive multimodal interactive memory assistant, empowering users to revisit, engage with, and manage their personal memories with greater depth and control.

\bibliographystyle{ACM-Reference-Format}
\bibliography{reference}

\newpage

\appendix
\setcounter{figure}{0}

\section{Prompts for LLMs}
\label{app:prompt}

\subsection{Identifying Composite Contexts}
\label{app:prompt_context}
\begin{lstlisting}
System instruction:
You are an intelligent agent capable of generating a list of COMPOSITE CONTEXTS inferred from the given memory. 
Composite context refers to a combination of time, location, people, objects, environment and activities. Such composite contexts could be inferred from the explicit content (e.g., text showing the event info) or implicit cues (e.g., multiple changes in location indicating travel). Focus on relatively important composites such as travel, conferences, and important meetings and focus less on trivial events. 
For each composite context, identify the related episodic memory ids. This could be due to time (e.g., the memory occurs during the event), location (e.g., the memory takes place at the event location), or specific content (e.g., the memory mentions the event).
Additionally, rate the importance of each event on a scale from 1 to 3, where 3 denotes very major events (e.g., multi-day events or highly important events), 2 denotes moderately important events, and 1 denotes less important events.

Exemplar composite context types include:
An academic conference: "An academic conference";
Recreational travel: "Trip to Salt lake city", "Traveling to home town";
Locational change: "Location changed from Seattle to Irvine";
Outdoor activities: "Camping trip";
Personal milestones: "Birthday celebration", "Graduation ceremony", "first day in univeristy";
etc.

Output the list of composite context in a JSON object with the key 'composite_context'. Each event should be represented as a sub JSON object with the following keys: 'event_name' (detailed and concise), 'memory_ids' (list), 'start_date', 'end_date' (could be the same as start_date), 'location', 'is_multi_days', and 'importance'.
+ 
<List of structured captured memories>
\end{lstlisting}

\subsection{Inferring Semantic Knowledge}
\label{app:prompt_knowledge}

\begin{lstlisting}
System instruction:
You are an intelligent agent capable of generating a list of FACTS or KNOWLEDGE (referred to knowledge in the following) that can be inferred from the given memory and the related composite contexts. Focus on relatively important high-level semantic knowledge and focus less on trivial events. Avoid specific details about individual media
The knowledge should be detailed and self-contained.
Exemplar semantic knowledge includes:
<Examples of semantic knowledge>
Also identify the most representative episodic memories that contribute to the understanding of the knowledge.
Output a JSON object with the key 'knowledge'. Each knowledge item should include 'knowledge', 'memory_ids' (list)

Input:
<Structured captured memories in the sliding windows>
+
<Identified composite contexts identified in the sliding window>
\end{lstlisting}

\begin{colortext}
\subsection{Query Augmentation}
\label{app:prompt_query_aug}

\begin{lstlisting}
System instruction:
augment_query = Given a query and today's date, identify the contextual filters. Contextual filters may include: 
temporal information: e.g., "last week."
location information: e.g., "Hawaii."
visible objects: e.g., "poke bowl."
people Seen: e.g., "people at the conference."
activities performed: e.g., "ordering in a restaurant."
and more complex contexts such as events or travel which consists of multiple atomic contexts mentioned above: e.g., "traveling to Hawaii." (activity: travel, location: Hawaii).
The query may not contain detailed contextual filters. In such cases, make reasonable inferences. For example, for query "What products did I buy from Sephora", the result could be obtained from a Sephora receipt. Thus inferred contextual filters for objects might be "makeup/skincare products or receipts."

Output a JSON object with the key 'augmented_query', including the sub-keys 'start_date', 'end_date', 'location', 'objects', 'people', 'activities', and 'complex_context'. Each sub-key should be a single string. Leave any sub-key empty if not applicable.
\end{lstlisting}
\end{colortext}

\subsection{Generating Answers Based on Retrieved Results}
\label{app:prompt_answer}

\begin{lstlisting}
System instruction:
Given a query, a list of memories and personal knowledge, generate a comprehensive answer to the query. 
Identify the episodic memories that can provide evidence to the question. 
If the answer is not explicitly presented in the memories, make a reasonable inference. 
Output a JSON object with the key 'answer', 'explanation' and 'memory_ids'. 
The 'answer' should be a string and 'memory_ids' should be a list of memory ids

Input:
<Query>
+
<Retrieved semantic knowledge>
+
<Retrieved structured knowledge>
\end{lstlisting}

\section{Baseline Implementation}
\label{app:baseline}

While there is no already-existing system designed for answering personal questions on captured memories, we manually designed and implemented a system as the baseline.
Similar to \codename, the baseline system also adopts an RAG architecture to adapt to the large number of captured memories.
We utilized the basic structure of RAG illustrated in~\cite{gao2023retrieval}, which involves (1) indexing the external data sources with embedding models, (2) leverage vector-based search to retrieve the top K relevant data instances (3) based-on the retrieved data, utilizing a powerful LLM to generate the final answer.
Note that typical RAG systems require a chunking phase, where long documents are split into smaller chunks for more precise matching and retrieval of relevant information.
In our case, each captured memory already represents a limited amount of information and is naturally separated.
Therefore, we treat each captured memory as an individual chunk.

\begin{figure*}[h]
    \centering
    \includegraphics[width=1\linewidth]{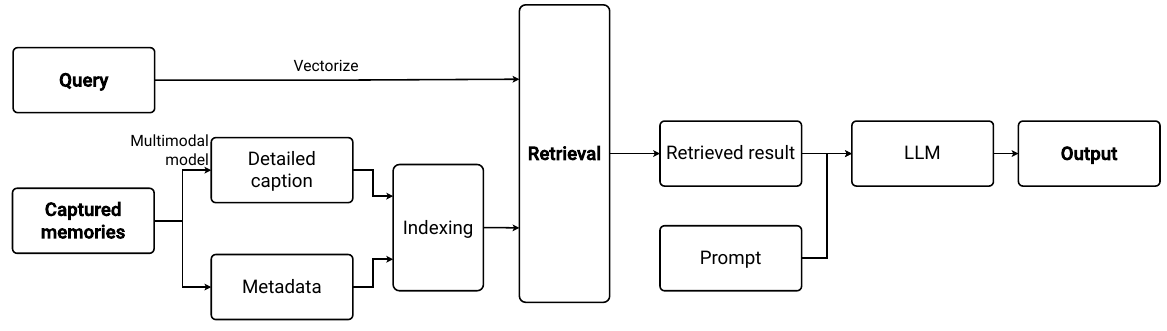}
    \caption{Structure of the baseline implementation.}
    \label{fig:baseline}
\end{figure*}

Figure \ref{app:baseline}\ref{fig:baseline} demonstrates the structure of the baseline system in our experiment.
The baseline also processes the captured memories by leveraging a multimodal model (\texttt{GPT-4o}) to generate detailed captions for each memory. 
Additionally, it extracts temporal and geographical information from the metadata and processes it in the same manner as \codename. 
This ensures that the processed memories include the temporal and geographical data, which are common components in users' queries.
The temporal and geographical information is concatenated to the generated caption.
Then the concatenated text sequence is encoded into text embeddings using embedding models (\texttt{text-embedding-3-small}). 

In the retrieval stage, the query is first encoded into the text embeddings using the same embedding model, and then retrieve the top K (K=50) captured memories using vector-based similarity search.
The retrieved top K captured memories are then ordered in temporal sequence, and then sent to the LLM (\texttt{GPT-4o)} for generating the answer. 
The prompt used for the answer generation is the same as \codename.

\begin{colortext}
\section{Failure Cases Analysis}
\label{app:failure_case}

\subsection{Failure Case Categorization}
Analyzing failure cases is important to understanding the limitations and improving the design of \codename.
Among the 137 queries tested in the study, we identified 25 queries with inaccurate results (\codename's~UPA~$\leq$~2) as failure cases.
Additionally, we reached out to participants asking them to manually retrieve the correct memories for these failures.
Through this analysis, we categorized them and propose future solutions for each:

\noindent\textbf{Case 1: Ambiguity (8 cases):}
Ambiguity in language-based interaction was the cause of failure in certain cases based on our analysis. Specifically, such ambiguities can be categorized as follows:

\begin{enumerate}[leftmargin=4.5mm]
    \item \textbf{Wording ambiguity}:
    P3 asked "what was the pool place I went in NYC". While they were referring to a billiards place (less ambiguous term), \codename interpreted it as a swimming pool, resulting in retrieval failure (Figure~\ref{fig:failure_cases}a). 

    \item \textbf{Reference ambiguity due to multiple valid answers}:
    Some queries have multiple potential answers.
    For example, P4 asked "What is the price of the medicine I bought?"
    They were referring to the most recent hospital visit, but \codename retrieved a different medicine receipt from the memory, leading to failure in answering users' question.

    \item \textbf{Contextual ambiguity:}
    P6, a photographer, asked, "How many times did I work as a photographer in the past few months?" 
    As they also enjoy personal photography, the boundary between photos taken for work and those taken for leisure is ambiguous, causing the system to fail.
    
\end{enumerate}
The above presents the challenges \codename faces in addressing ambiguity, both in understanding user queries and in analyzing retrieved results. For a detailed discussion on strategies to address these uncertainties, please refer to Section \ref{sec:discussion_uncertainty}.


\noindent\textbf{Case 2: Lack of contextual cues (7 cases):}
Several failure cases occurred due to insufficient contextual information to associate the target memory with the input query. There are two types:
\begin{enumerate}[leftmargin=4.5mm]
    \item \textbf{Target context being too implicit:}
    As shown in Figure~\ref{fig:failure_cases}b, P4 asked ``What is the content of the last meeting with my advisor last week?''
    The target memory, a photo of a notebook page, lacks contextual cues to associate it with the meeting.
    This led to retrieval failure.
    In such cases, \codename should ask the user to clarify or iterate on the query (e.g., specifying the type of memory if the user has an idea).

    \item \textbf{People Identity and Metadata:}
    For example, P5 asked ``Where did I travel with two Korean friends last month?''
    \codename currently lacks access to facial recognition or metadata that can identify and associate individuals by attributes such as race. This also led to retrieval failure when answering this question. 
    Future work could integrate with platforms like Google Photos or Apple Albums, which group photos by individuals using facial recognition. Additionally, users could manually add metadata via linking photos to contacts or descriptive tags (e.g., ``friends from summer school''), enabling the system to handle such queries better.
\end{enumerate}

\begin{figure}
    \centering
    \includegraphics[width=1\linewidth]{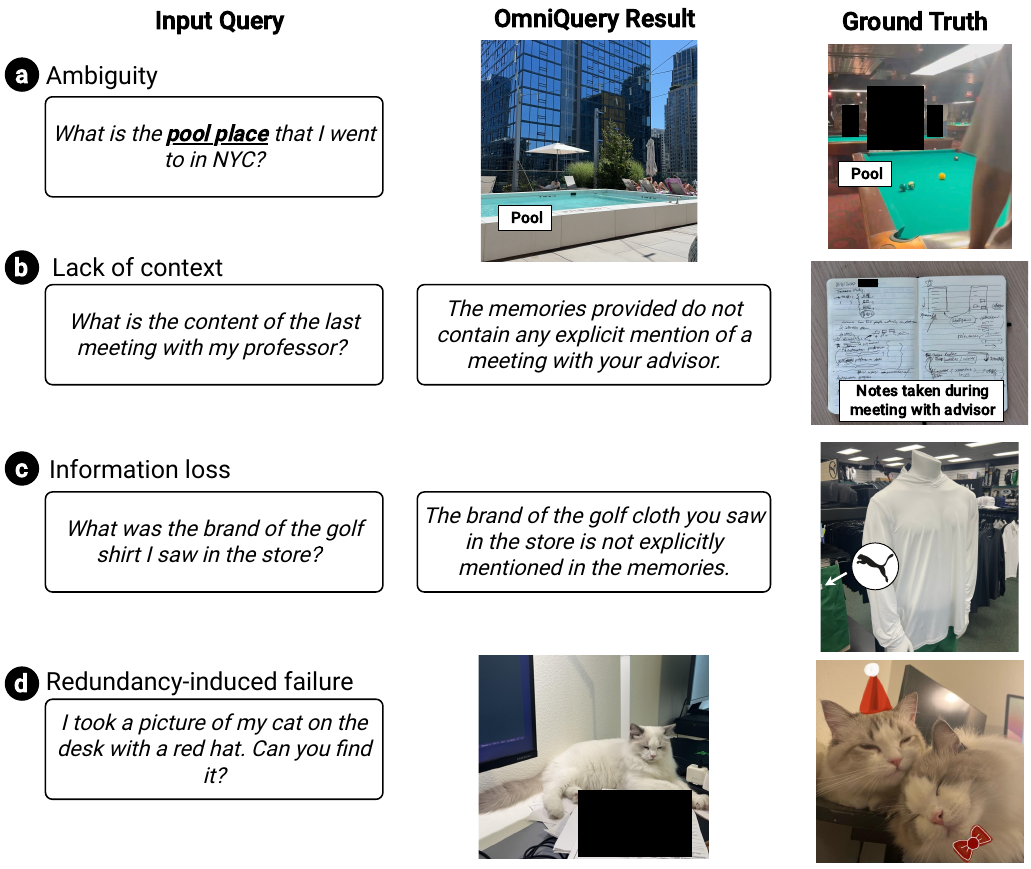}
    \caption{\begin{colortext}Four exemplar failure cases: (a) lack of context, (b) wording ambiguity, (c) information loss during processing and (d) redundancy-induced failure.\end{colortext}}
    \label{fig:failure_cases}
\end{figure}


    

\noindent\textbf{Case 3: Information loss during text-based preprocessing (3 cases):}
\codename currently adopts a text-based augmentation to extract atomic, composite context and semantic knowledge, which might lead to information loss.
For example, P10 asked ``What was the brand of the golf shirt I saw in the store?''
The brand logo was barely visible in the bottom-left corner in the target memory (Figure~\ref{fig:failure_cases}c) and was not captured during the preprocessing.
This led to retrieval failures when answering the question.
Future work could integrate current text-based retrieval with advanced multimodal retrieval models (\eg ColPali~\cite{faysse2024colpaliefficientdocumentretrieval}), which are capable of keeping more details during the retrieval process.



\noindent\textbf{Case 4: Redundacny-induced failure (3 cases):}
Similar to the ``needle in a haystack'' challenge~\cite{li2024needlebenchllmsretrievalreasoning}, \codename's performance might degrade when too many memories are retrieved during the retrieval phase.
For example, P7 frequently takes photos of their cats and when they asked ``I took a picture of my cat on the desk with a red hat. Can you find it?''
\codename tries to retrieve all relevant memories about their cats, resulting in failure to find the correct one~(Figure~\ref{fig:failure_cases}d). 
In contrast, as the baseline system always retrieves a fixed number of results, it is able to identify the correct answer by narrowing down the search space.
To address this, query-aware filtering process such as reranking~\cite{dong2024don} could be employed to narrow down the search space.
Additionally, employing a Top-K retrieval strategy~\cite{khattab2020finding} could provide users with more options to enhance overall performance.



\noindent\textbf{Case 5: Subjectivity-induced failure (1 case):}
P2 asked, ``Give me the best selfie I took.''
However, the subjective nature of ``the best'' made it difficult for \codename to determine the correct answer. 
To address such cases, future systems could integrate user preferences (e.g., leveraging marked favorites) or enable interactive clarifications (e.g., asking, ``Do you prefer an indoor or outdoor selfie?'') to better align with user intent.

\noindent\textbf{Case 6: Target memory out of scope (3 cases):}
In three cases, the target memory was unavailable.
In one case, the target memory was outside the 100-file range while the participants thought it had been included. In the other two cases, participants mistakenly thought they had captured the memory, but it was not actually captured.
It is important to communicate such uncertainties to users when the system believes a query to be not answerable. For detailed discussion, please refer to Sec \ref{sec:discussion_uncertainty}.

\subsection{Perceived Feelings from Participants} We also present cases when participants reacted negatively to the answers. 
All participants encountered cases where the answers are inaccurate.
Some were incomplete (e.g., P1 believed that they visited mroe than a few churches on the trip to Barcelona, but answers provided only two of them). Some were presumptive (e.g., P7 asked about recent social events, where the answers gave a piece of memory on a museum visit and explained that visiting museums is ``likely with other people.'' However, P7 visited the museum alone). Some were making mistakes (e.g., P7 asked for the mostly visited attractions but both system mistakenly answered a museum, which was because P7 took a lot of museum pictures and both systems failed to recognize that they were the same visit.) Some even more challenging questions that caused failure of both systems include questions relate to a specific person. For example, P8 asked about her significant other and P5 asked about their ``Korean friend'' met in a trip. These cases represent the difficulty of understanding the nuances of personal relationships with personal album data.

\subsection{Cases Where the Baseline RAG Outperforms \codename}
As discussed above, in cases where there is redundancy in the retrieved results before generating the answer, the baseline RAG system may perform better than \codename because it retrieves a fixed number of memories, narrowing the input context and reducing noise during answer generation.
To further understand the comparison, we also examined cases where the baseline performs better even when both results are reasonably accurate (UPA $\geq$ 3).
Typically, in cases where there is ambiguity in the query, while \codename might provide a relatively accurate result, the baseline often produces a more binary outcome (either highly accurate or highly inaccurate) due to its narrower retrieval scope. This reduces ambiguity but also limits its ability to handle complex contexts.




\end{colortext}

\begin{colortext}
    
\section{Curating a Fixed Dataset Discussion}
\label{app:fixed_dataset}

\codename is evaluated through real user data, and its effectiveness can be further evaluated on a fixed benchmark dataset. However, curating a fixed benchmark dataset for personal data presents significant challenges:
\begin{itemize}[leftmargin=3mm]
    \item \textbf{Difficulties in collection of long-term personal data while preserving privacy}:
    Collecting diverse, long-term personal data while preserving participants' privacy with proper consent and redaction is complex.
    Prior work like Ego4D ensures privacy by obtaining consent for controlled indoor environments or by de-identifying data through redaction of visible and audible PII~\cite{Grauman2021Ego4DAT}. 
    However, personal captured memories inherently include interactions with various people and sensitive personal content (e.g., photos of IDs or financial documents), which makes it impractical to obtain universal consent or redact all PII without compromising the data's utility. 
    A promising solution is advanced generative content replacement (e.g., Xu \etal~\cite{xu2024chi}), which replaces sensitive PII with synthetic content, ensuring privacy while preserving the cabality of benchmarking.

    \item \textbf{Difficulties in generating objective and unbiased QA pairs}:
    Personal captured memories and corresponding questions are by definition subjective and sometimes ambiguous.
    Subjectivity varies across question types. Some focus on objective facts (e.g., ``What is the Wifi password?''), while others could be highly subjective (e.g., ``What sports were my favorite last year?''). 
    To address this, research methods can be applied to reduce ambiguity and bias, such as leveraging crowdsourcing to create QA pairs from third-person perspectives or assess the objectivity of existing pairs. 
    Future work should aim to develop a taxonomy of question types and explore strategies for guiding crowd workers to assess objectivity or generate improved QA pairs based on the taxonomy.

\end{itemize}
\end{colortext}

\end{document}